\documentclass[aps,twocolumn,floatfix,superscriptaddress]{revtex4-1}

\usepackage[british]{babel}
\usepackage[utf8]{inputenc}
\usepackage{lmodern}
\usepackage[T1]{fontenc}
\usepackage{mathtools} 
\usepackage{tikz}
\usetikzlibrary{decorations.markings}
\usepackage{epsfig}
\usepackage{float}
\usepackage{placeins}
\usepackage{hyperref}
\usepackage{booktabs} 
\usepackage{tabularx} 
\usepackage{multirow} 

\begin{document}

\title{Finite-size scaling at infinite-order phase transitions}
\author{Rick Keesman}\affiliation{Instituut-Lorentz,
  Leiden University, Niels Bohrweg 2, 2333 CA Leiden, The
  Netherlands}

\author{Jules Lamers}\affiliation{Institute for Theoretical Physics and Center for Extreme Matter and Emergent Phenomena, Utrecht University, Leuvenlaan 4, 3584 CE Utrecht, The Netherlands}

\author{R.~A.~Duine}\affiliation{Institute for Theoretical Physics and Center for Extreme Matter and Emergent Phenomena, Utrecht University, Leuvenlaan 4, 3584 CE Utrecht, The Netherlands}\affiliation{Department of Applied Physics, Eindhoven University of Technology, PO Box 513, 5600 MB, Eindhoven, The Netherlands}

\author{G.~T.~Barkema} \affiliation{Instituut-Lorentz,
  Leiden University, Niels Bohrweg 2, 2333 CA Leiden, The
  Netherlands}\affiliation{Institute for Theoretical Physics and Center for Extreme Matter and Emergent Phenomena, Utrecht University, Leuvenlaan 4, 3584 CE Utrecht, The Netherlands}

\begin{abstract}
	\noindent For systems with infinite-order phase transitions, in which an order parameter smoothly becomes nonzero, a new observable for finite-size scaling analysis is suggested. By construction this new observable has the favourable property of diverging at the critical point. Focussing on the example of the \textit{F}-model we compare the analysis of this observable with that of another observable, which is also derived from the order parameter but does not diverge, as well as that of the associated susceptibility. We discuss the difficulties that arise in the finite-size scaling analysis of such systems. In particular we show that one may reach incorrect conclusions from large-system size extrapolations of observables that are not known to diverge at the critical point. Our work suggests that one should base finite-size scaling analyses for infinite-order phase transitions only on observables that are guaranteed to diverge.
\end{abstract}

\pacs{pacs}

\maketitle

\section{Introduction}\label{sec1}

\noindent The study of phase transitions is a central topic in physics. In statistical physics these drastic changes in the physical properties of a system show up in non-analytic behaviour of quantities such as the free energy~$f$ per volume. For finite-order phase transitions (FOPTs) this takes the form of non-smoothness, where some derivative of $f$ makes a jump at the critical temperature. Such discontinuous functions provide suitable observables for numerical investigation into universal as well as model-specific properties of the phase transition. In this setting finite-size scaling (FSS) is a powerful tool to quantitatively extrapolate the power-law behaviour of observables near criticality~\cite{FF_69,NB_10}.

For infinite-order phase transitions (IOPTs) the situation is more subtle since the transition is not as abrupt as for FOPTs. In the prototypical example, the \textit{XY}-model, the critical---or perhaps more appropriately `transition'---temperature marks the point at which free vortices start to dominate the physics, even though the susceptibility, which characterizes the single-vortex fluctuations, has a peak away from this temperature \cite{Min_87}.
From a more mathematical perspective the non-analyticity marking IOPTs is rather weak: the free energy depends smoothly on the temperature, where $f$ and all its derivatives are continuous, but it has an essential singularity at the critical temperature. (Recall that, unlike in the complex case, there are smooth functions that are not real-analytic; a standard example is the function given by $\exp(-1/x)$ for $x>0$ and zero elsewhere.) In addition IOPTs often exhibit logarithmic finite-size corrections~\cite{Bax_07,Bar_83,San_13}; although this does not make FSS impossible~\cite{HB_80,*HB_81} it has been shown to give rise to difficulties~\cite{Ken_05}, and rather large systems must be investigated to accurately analyse the scaling. Accordingly, various other numerical methods for studying IOPTs have also been developed~\cite{Swe_77,HDB_11,BN_93}.

In such a more delicate setting one has to take care to select appropriate observables for numerical analysis using FSS. Order parameters do not directly allow one to locate the critical point  for IOPTs since the numerical determination of the point at which a function smoothly becomes nonzero is a futile task. For this reason observables that diverge at the critical point, e.g.~susceptibilities for second-order phase transitions, are more suitable for studying a model's behaviour near criticality~\cite{Maz_92,WJ_05,Ken_05}. One should also keep in mind that for IOPTs there are also observables, such as the specific heat, that do not diverge for increasing system size; they peak away from the critical temperature and do not tend to a Dirac delta function in the thermodynamic limit of infinite system size~\cite{Bar_83}. In this work we propose a new observable that, by construction, peaks at the critical temperature in the thermodynamic limit for any model with an IOPT that is characterized by a smooth order parameter.

Specifically we consider the \textit{F}-model, which is an interesting test case since it was solved analytically on a square lattice with periodic boundaries in the thermodynamic limit~\cite{Lie_67a,Lie_67b}. At the same time it is related to the \textit{XY}-model via a series of dualities involving the discrete Gaussian solid-on-solid model and the Coulomb gas~\cite{Lie_67a,Bei_77,Nie_83,Sav_80}. Our new observable is essentially the logarithmic derivative of the spontaneous staggered polarization~$P_0$, for which an asymptotic analytical expression is known for all temperatures~\cite{Bax_73b}. We use a FSS analysis to compare the new observable with the ordinary derivative of $P_0$ and the susceptibility associated with $P_0$. These observables behave quite differently: the logarithmic derivative nicely diverges at the critical point in the thermodynamic limit, the ordinary derivative has a bounded peak elsewhere for all system sizes, and for the susceptibility---which is commonly used to analyse critical behaviour---the scaling near criticality in the thermodynamic limit has been conjectured~\cite{Bax_73a}. In our estimates of characteristics such as the critical temperature, however, identical analyses of these observables lead to similar asymptotic results. This once more illustrates that one should be careful in numerical analyses of IOPTs. In particular, our work thus suggests that one should base FSS analyses for IOPTs only on observables that are guaranteed to diverge.

This paper is organised as follows. In Section~\ref{sec:sec2} we recall the basics of the \textit{F}-model and discuss the relevant observables and their known asymptotic expression. The Monte Carlo cluster algorithm and data processing are treated in Section~\ref{sec:sec3}. The analysis of the three observables is performed in Section~\ref{sec:sec4}, and the results are discussed in Section~\ref{sec:sec5}. We end with a conclusion in Section~\ref{sec:sec6}.

\section{The \textit{F}-model and observables}\label{sec:sec2}

\begin{figure}[!ht]
	\centering
	\begin{tikzpicture}[decoration={markings, mark=at position 0.6 with {\arrow[scale=1.5,>=stealth]{>}}},font=\normalsize,scale=2]
	
	\draw[postaction=decorate] (0,.5) -- (.5,.5);
	\draw[postaction=decorate] (.5,.5) -- (1,.5);
	\draw[postaction=decorate] (.5,0) node[below] {$\epsilon_1=\epsilon_{\hphantom{1}}$} -- (.5,.5);
	\draw[postaction=decorate] (.5,.5) -- (.5,1);
	
	\begin{scope}[shift={(1.5,0)}]
	\draw[postaction=decorate] (0,.5) -- (.5,.5);
	\draw[postaction=decorate] (.5,.5) -- (1,.5);
	\draw[postaction=decorate] (.5,.5) -- (.5,0) node[below] {$\epsilon_3=\epsilon_{\hphantom{1}}$};
	\draw[postaction=decorate] (.5,1) -- (.5,.5);
	\end{scope}
	
	\begin{scope}[shift={(3,0)}]
	\draw[postaction=decorate] (0,.5) -- (.5,.5);
	\draw[postaction=decorate] (1,.5) -- (.5,.5);
	\draw[postaction=decorate] (.5,.5) -- (.5,0) node[below] {$\epsilon_5=0\ \, $};
	\draw[postaction=decorate] (.5,.5) -- (.5,1);
	\end{scope}
	
	\begin{scope}[shift={(0,-1.5)}]
	\draw[postaction=decorate] (.5,.5) -- (0,.5);
	\draw[postaction=decorate] (1,.5) -- (.5,.5);
	\draw[postaction=decorate] (.5,.5) -- (.5,0) node[below] {$\epsilon_2=\epsilon_{\hphantom{1}}$};
	\draw[postaction=decorate] (.5,1) -- (.5,.5);
	\end{scope}
	
	\begin{scope}[shift={(1.5,-1.5)}]	
	\draw[postaction=decorate] (.5,.5) -- (0,.5);
	\draw[postaction=decorate] (1,.5) -- (.5,.5);
	\draw[postaction=decorate] (.5,0) node[below] {$\epsilon_4=\epsilon_{\hphantom{1}}$} -- (.5,.5);
	\draw[postaction=decorate] (.5,.5) -- (.5,1);
	\end{scope}
	
	\begin{scope}[shift={(3,-1.5)}]			
	\draw[postaction=decorate] (.5,.5) -- (0,.5);
	\draw[postaction=decorate] (.5,.5) -- (1,.5);
	\draw[postaction=decorate] (.5,0) node[below] {$\epsilon_6=0\ \, $} -- (.5,.5);
	\draw[postaction=decorate] (.5,1) -- (.5,.5);
	\end{scope}
	\end{tikzpicture}
	
	\caption{The six allowed vertices with associated energies for the \textit{F}-model, where $\epsilon>0$.}
	\label{fig:sixvertices}
\end{figure}
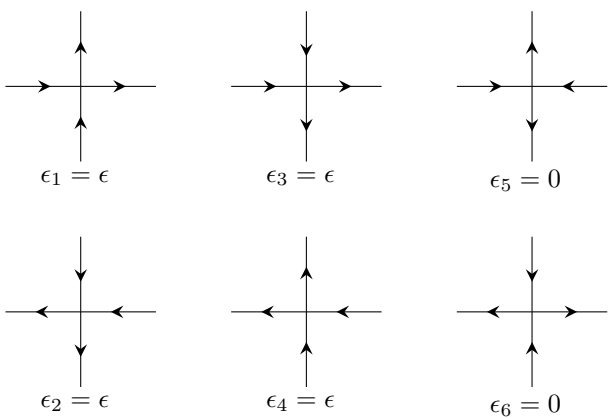

\noindent The six-vertex model, or ice-type model, is a lattice model for which each vertex is connected to four others by edges carrying an arrow pointing in or out of the vertex, such that precisely two arrows point towards each vertex. Thus there are six allowed configurations around each vertex as shown in Figure~\ref{fig:sixvertices}. To each such vertex configuration~$i$ one assigns a (local) Boltzmann weight~$\exp({-\beta} \, \epsilon_i)$, where $\beta\coloneqq1/(k_\text{B} T)$ is the inverse temperature and $\epsilon_i$ the energy of that configuration. The (global) Boltzmann weight of the entire configuration is the product of the local weights of all vertex configurations. The \textit{F}-model~\cite{Rys_63} is given by the particular choice $\epsilon_1=\epsilon_2=\epsilon_3=\epsilon_4=\epsilon>0$ and $\epsilon_5=\epsilon_6=0$. This is the prototype of the antiferroelectric regime of the six-vertex model, where vertex configurations 5 and~6 are energetically favourable. At sufficiently low temperatures the system orders in an antiferroelectric fashion, with vertices 5 and~6 alternating in a checkerboard-like fashion. From now on we consider the \textit{F}-model on a square $L\times L$ lattice with periodic boundary conditions in both directions, and set $k_\text{B} = \epsilon = 1$.

The free energy (per site) in the thermodynamic limit was found analytically for the \textit{F}-model by Lieb~\cite{Lie_67a,Lie_67b} using a Bethe-ansatz analysis. There is an IOPT with critical (or `transition') temperature $\beta_\text{c}=\ln2$, or $\Delta_\text{c}=-1$ where $\Delta \coloneqq 1-\exp(2 \beta)/2$. In the low-temperature regime the free energy can be expressed as a convergent series,
\begin{align}\label{eq:fanalow}
\beta \, f^{\text{ana}}(\lambda) =
    \beta - \lambda - \sum_{n=1}^{\infty} \frac{\exp({-n} \lambda) \sinh(n \lambda)}{n \cosh(n \lambda)}
\end{align}
where $\lambda \coloneqq \text{arccosh}({-\Delta})>0$ parametrizes $\beta>\beta_\text{c}$, while at high temperatures one has an integral representation
\begin{align}\label{eq:fanahigh}
&\beta \, f^{\text{ana}}(\mu) = \beta\, - \\
    & \qquad\qquad \frac{1}{4 \mu} \int_0^{\infty} \! \frac{dt}{\cosh(\pi t/2\mu)} \ln\!\left( \frac{\cosh(t) - \cos(2 \mu)}{\cosh(t) - 1} \right) \nonumber
\end{align}
for $\mu \coloneqq \arccos({-\Delta})$, $0<\mu<\pi/2$, parametrizing $\beta<\beta_\text{c}$. The entire high-temperature region can be regarded as critical in the sense that correlations decay as inverse power laws rather than exponentially~\cite{Bax_07}.

Although the six-vertex model has not been solved in the presence of an external staggered electric field, Baxter~\cite{Bax_73b} found an exact expression for the spontaneous staggered polarization $P_0$ per site. To each microstate~$C$ one can associate an `instantaneous' spontaneous staggered polarization $P_0(C)$, which can be computed as the `staggered' sum of the net polarizations at the vertices, where the direction of the net polarization is flipped at every other site (in a checkerboard-like way). Then the thermal average $P_0 \coloneqq \langle P_0(C) \rangle$ is an order parameter for the \textit{F}-model, vanishing for $\beta<\beta_\text{c}$ and becoming nonzero at the critical temperature. When $\beta>\beta_\text{c}$ it is given by
\begin{equation}\label{eq:panalow}
	P_0^\text{ana}(\lambda)^{1/2} = \sqrt{\frac{2 \pi}{\lambda}} \, \sum_{n=1}^{\infty} \exp\!\left({-\frac{(n-1/2)^2 \pi^2}{2\lambda}}\right).
\end{equation}
Like the free energy this function is smooth with an essential singularity at $\beta=\beta_\text{c}$, which is very weak: the functions and all their derivatives do tend to zero as $\beta$ approaches $\beta_\text{c}$ from above. When the \textit{F}-model is reinterpreted as a height model (the body-centred solid-on-solid model) the IOPT is a roughening transition~\cite{Bei_77}.

The observables on which we will focus are the derivatives $\beta^2 \, \partial_\beta \ln P_0$ and $\beta^2 \, \partial_\beta P_0$, where $\partial_\beta\coloneqq \partial/\partial\beta$, together with the susceptibility $\chi \coloneqq \beta \, [\langle P_0(C)^2 \rangle - \langle P_0(C) \rangle^2]$ of the staggered polarization, which is called the spontaneous staggered polarizability. Baxter~\cite{Bax_73a} conjectured the following form of the susceptibility in the low-temperature regime
\begin{equation}\label{eq:chicon}
	\chi(\lambda) \sim \lambda^{-2} \exp(\pi^2/2 \lambda) \, .
\end{equation}
\indent The preceding discussion ensures that $\beta^2 \, \partial_\beta \ln P_0$ diverges at the critical temperature whereas $\beta^2 \, \partial_\beta P_0$ has a (finite) peak at some $\beta_\text{max}>\beta_\text{c}$. To the best of our knowledge neither $\beta^2 \, \partial_\beta \ln P_0$ nor $\beta^2 \, \partial_\beta P_0$ have been considered before in the literature. The latter is included to demonstrate one has to be careful in FSS for IOPT: we show that it is hard to extrapolate numerical data to the thermodynamic limit, even when the exact limiting expressions are known.

\section{Simulations}\label{sec:sec3}

\noindent Our Monte Carlo simulations are based on a cluster algorithm that uses the (one-to-three) mapping from the six-vertex model to a three-colouring of the square lattice~\cite[Note added in proof]{Lie_67a}. Choose three colours, ordered in some way, and use one of them to colour any single plaquette (face) of the lattice. Then any configuration of the six-vertex model uniquely determines a three-colouring, where the direction of the arrow on an edge dictates whether the colour increases or decreases (modulo three), and the ice rule ensures that the colouring is well defined. For the \textit{F}-model vertices surrounded by all three colours (configurations 1 to~4 in Figure~\ref{fig:sixvertices}) are energetically less favourable than those at which only two colours meet (configurations 5 and~6).

The multi-cluster algorithm builds clusters containing adjacent faces of two colours, and patches these clusters together diagonally with a probability that is such that required detailed balance is met. After no more clusters can be included the colours in the clusters are swapped and one cluster update has been performed~\cite{NB_10}. Because of the small auto\-correlation times at the temperatures near the phase transition, we take measurements after $10$ of these cluster updates for system sizes $L<128$, and after each cluster update for larger systems. At least $10^6$ measurements are made per temperature per system, at minimally $15$ different temperatures. For the largest system that we consider, with $L=512$, we simulate at $29$ different temperatures with slightly over $8\times10^6$ measurements performed per temperature.

From expressions \eqref{eq:fanalow}--\eqref{eq:fanahigh} for the free energy we can estimate the mean and variance in energy measurements for finite systems at a given temperature. Moreover the specific heat $C_v=\beta^2 \partial_\beta^2 (\beta f)$ is bounded and, in leading order, does not scale with $L$. Together these ensure that the parallel-tempering and multi-histogram methods can be applied successfully.

Parallel tempering is a simulation method in which systems are simulated at various temperatures and periodically swapped \cite{Par_92}. Here the probability of swapping two configurations at different temperatures is given by $P_{\text{swap}}=\min[1,\exp(\delta_\beta \, \delta_E)]$, where $\delta_\beta \coloneqq \beta_{\text{high}}-\beta_{\text{low}}$ and $\delta_E \coloneqq E_{\text{high}}-E_{\text{low}}$ are the difference in inverse temperature and energy between the two configurations, respectively. To make sure that $P_{\text{swap}}$ is large enough for configurations to move reasonably fast through this temperature landscape we want the histograms of the energies at different temperatures to overlap significantly. Starting from some temperature for which we know the average energy $U \coloneqq \langle E(C) \rangle$ and the standard deviation $\sigma_U$ from the analytical expression of the free energy, a neighbouring temperature is chosen such that the difference in energies is roughly $\sigma_U$, viz.~$\beta' = \beta \pm \beta/\sqrt{C_v}$. After each measurement we may swap the configuration with one at such a neighbouring temperature, with acceptance probability~$P_\text{swap}$ between $47\%$ and $53\%$ for all simulations at large system sizes.

\begin{figure}[ht]
	\centering
	\epsfig{file=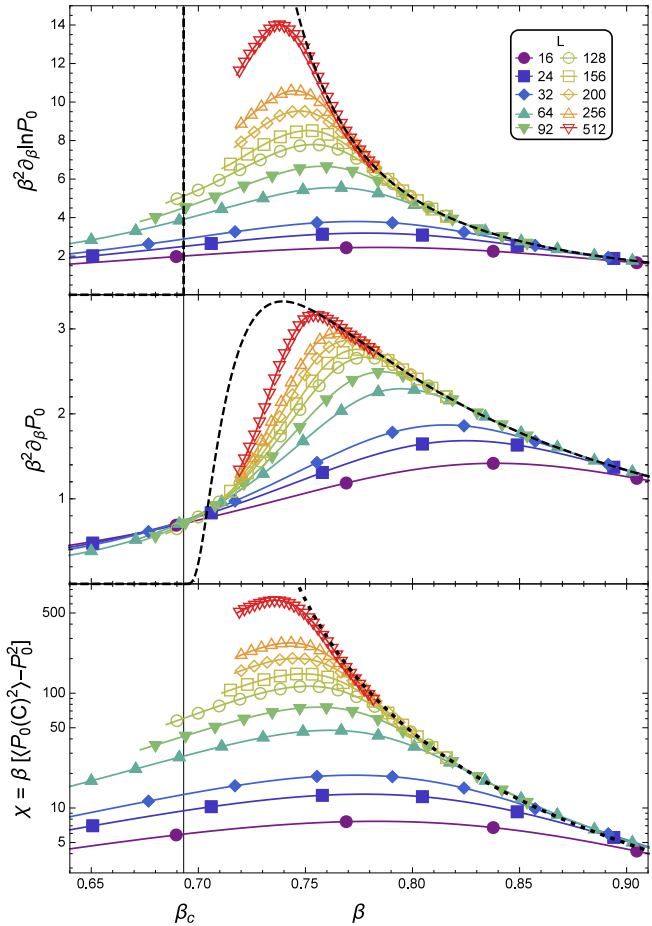,width=0.99\linewidth,clip=}
	\caption{The observables $\beta^2 \, \partial_\beta \ln P_0$ (upper panel), $\beta^2 \, \partial_\beta P_0$ (central panel), and $\chi$ (lower panel) versus $\beta$ for system sizes up to $L=512$. The data points show the temperatures at which the simulations were run, while the solid lines are the functions extracted from this data using the multi-histogram method. When available the analytical form for infinite systems, cf.~Eq.~\eqref{eq:panalow}, is shown by a dashed black line. For sufficiently low temperatures all graphs collapse onto these dashed black lines, corroborating the validity of our simulations. For $L\to\infty$ we know that $\beta^2 \, \partial_\beta \ln P_0$ must diverge at the critical temperature $\beta_\text{c}=\ln 2$, indicated by a vertical line, whereas $\beta^2 \, \partial_\beta P_0$ is bounded and peaks elsewhere. A fit to the conjectured form of $\chi$, Eq.~\eqref{eq:chicon}, is indicated by a dotted black line in the lower panel.}
    \label{fig:comparisonplot}
\end{figure}

At each measurement we record the energy~$E(C)$ and instantaneous spontaneous staggered polarization~$P_0(C)$ for various temperatures. Using the multi-histogram method any function of the values $E(C)$ and $P_0(C)$ can then be reliably estimated as a function of temperature~\cite{FS_89}. For this method to work the energy histograms must have significant overlap; we have ensured that this is indeed the case for our data. Figure~\ref{fig:comparisonplot} shows the result for $\beta^2 \, \partial_\beta \ln P_0$, $\beta^2 \, \partial_\beta P_0$ and $\chi$, together with their known and conjectured analytical form. Note that the data in the low-temperature regime are in agreement with the analytical forms of $\beta^2 \, \partial_\beta \ln P_0$ and $\beta^2 \, \partial_\beta P_0$. For $\chi$ the data collapse in this regime and support the conjecture~\eqref{eq:chicon}.

\section{Analysis}\label{sec:sec4}

\noindent The usual finite-size scaling procedure is to take the data, see Fig.~\ref{fig:comparisonplot}, and collapse the graphs by scaling the distance to the critical temperature and the height as functions of the system size~$L$. For the \textit{F}-model there are large logarithmic corrections due to `quasi' long-range correlations~\cite{Min_87} as well as higher-order finite-size corrections~\cite{WJ_05}. The systems size at which the finite-size corrections become negligible do not yet seem to be within reach, so we cannot perform a data collapse based purely on analytical expressions.

\begin{figure}[ht]
	\centering
	\epsfig{file=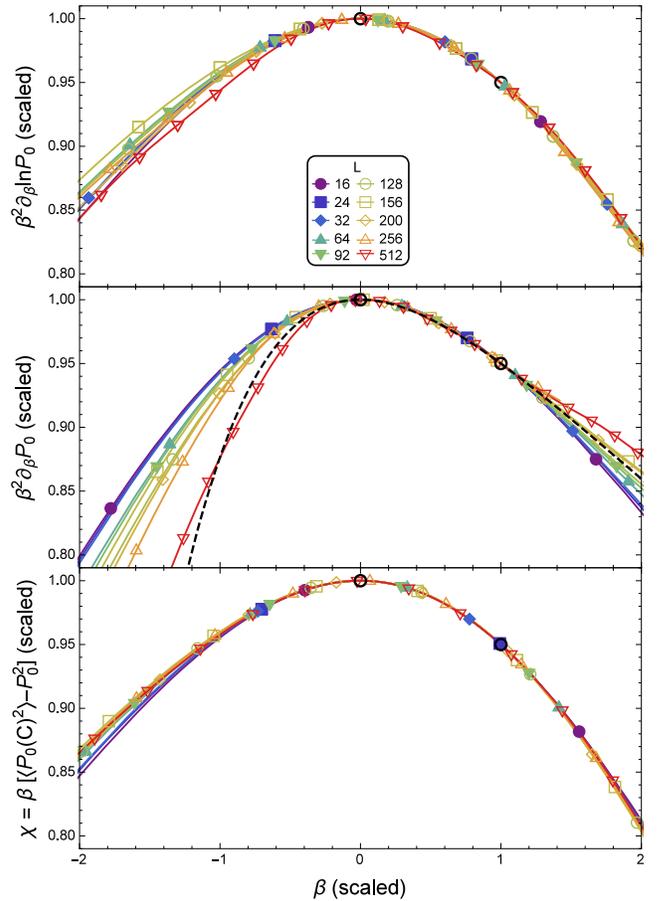,width=0.99\linewidth,clip=}
	\caption{The three observables scaled such that for each system size $(\beta_\text{max},h_\text{max})\mapsto(0,1)$ and $w\mapsto1$. This scaling works well in the low-temperature regime for $\beta^2 \, \partial_\beta \ln P_0$ (upper panel) and $\chi$ (lower panel). For $\beta^2 \, \partial_\beta P_0$ (central panel) it seems to fail, cf.~the deviation from the asymptotic analytical result indicated by a dashed black line.}
    \label{fig:collapseplot}
\end{figure}

Instead we will perform a numerical data collapse. For each of the three observables that we are interested in we determine the coordinates $(\beta_{\text{max}},h_{\text{max}})$ of the maximum, together with the peak width~$w$. Here we define the width by demanding that the function passes through the point $(\beta_\text{max}+w,0.95\,h_\text{max})$. This definition is chosen such that $w$ can be accurately measured for large systems given the simulation data; we focus on lower temperatures (higher~$\beta$) because of the asymmetry of the observables around the critical temperature. Thus we have three characteristics, which are well defined since any observable is smooth and bounded for finite systems. This allows for a numerical data collapse by shifting $(\beta_\text{max},h_\text{max})$ and $(\beta_\text{max}+w,0.95 \, h_\text{max})$ on top of each other. The result for our three observables is shown in Fig.~\ref{fig:collapseplot}.
Sufficiently close to the critical point $\beta^2 \, \partial_\beta \ln P_0$ and $\chi$ scale well, which is a positive sign for scalability to the thermodynamic limit. Note that $\beta^2 \, \partial_\beta P_0$, for which we know the (bounded) asymptotic solution, does not exhibit scalability for the system sizes that we investigate. We extrapolate the characteristics ($\beta_\text{max},h_\text{max})$ and $w$, extracted from the data for various system sizes, to the thermodynamic limit.

\subsection{Peak position~$\beta_\text{max}$}
\begin{figure}[h]
	\centering
	\epsfig{file=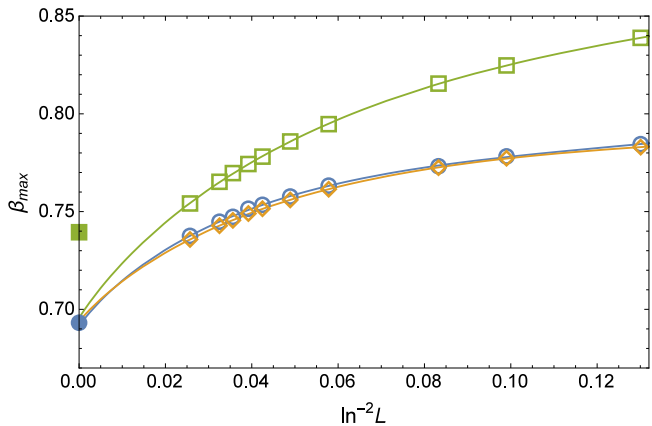,width=0.99\linewidth,clip=}
	\caption{The inverse temperatures at which $\beta^2 \, \partial_\beta \ln P_0$ (blue circles), $\beta^2 \, \partial_\beta P_0$ (green squares), and $\chi$ (yellow diamonds) are maximal, here shown as functions of the system size. The asymptotic solutions, $\beta_\text{c}=\ln 2$ for $\beta^2 \, \partial_\beta \ln P_0$ and $\beta_\text{max}^\text{ana}\approx0.7394$ for $\beta^2 \, \partial_\beta P_0$, are shown at $\ln^{-2} L=0$. Best fits of the form Eq.~\eqref{eq:bcritfit} to the data are shown as solid lines, and all seem to converge to $\beta_\text{c}$.}
    \label{fig:extra-bcrit}
\end{figure}

The analytic expression in Eq.~\eqref{eq:panalow} reveals that $\beta^2 \, \partial_\beta \ln P_0$ must develop a Dirac delta-like peak at $\beta_\text{c} \approx0.6931$ as $L\to\infty$. Instead, the peak of $\beta^2 \, \partial_\beta P_0$ remains finite and shifts to $\beta_\text{max}^\text{ana} \approx 0.7394$. The large-$L$ behaviour of the spontaneous staggered polarizability~$\chi$ is not analytically known. The form of the leading finite-size corrections can be obtained by expanding the inverse temperature in $L$ as~\cite{WJ_05}
\begin{align}\label{eq:bcritfit}
	 \beta_{\text{max}}(L)=\beta_\text{c} + \frac{A_\beta}{\ln^{2}L} + \frac{B_\beta}{\ln^{3}L} + \frac{C_\beta}{\ln^{4}L} \, .
\end{align}
Figure~\ref{fig:extra-bcrit} displays our results for $\beta_\text{max}$ as a function of $L$ as obtained from our three observables, together with the analytic asymptotic values, and the best fits to Eq.~\eqref{eq:bcritfit}. These fits yield $\beta_\text{max}^\text{fit}=0.6914(28)$ for $\beta^2 \, \partial_\beta \ln P_0$,
$\beta_\text{max}^\text{fit}=0.6955(17)$ for $\beta^2 \, \partial_\beta P_0$, and   $\beta_\text{max}^\text{fit}=0.6937(11)$ for $\chi$.

\subsection{Peak height~$h_\text{max}$}
\begin{figure}[h]
	\centering
	\epsfig{file=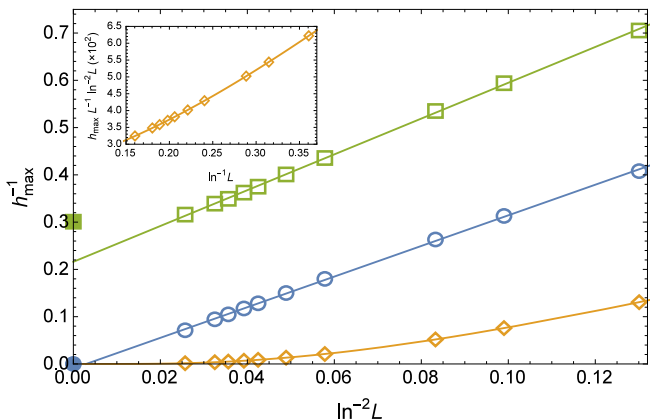,width=0.99\linewidth,clip=}
	\caption{The inverse maximal heights of $\beta^2 \, \partial_\beta \ln P_0$ (blue circles), $\beta^2 \, \partial_\beta P_0$ (green squares) and $\chi$ (yellow diamonds) as functions of $\ln^{-2} L$. The inset shows the peak height of $\chi$ with differently scaled axes. The asymptotic values, $(h_\text{max}^\text{ana})^{-1}=0$ for $\beta^2 \, \partial_\beta \ln P_0$ and $(h_\text{max}^\text{ana})^{-1}\approx0.3009$, are included at $\ln^{-2} L=0$. Data indeed suggests that $\beta^2 \, \partial_\beta \ln P_0$ and $\chi$ diverge while $\beta^2 \, \partial_\beta P_0$ stays finite. Best linear fits as functions of $\ln^{-2} L$ are shown as solid blue and green lines for $\beta^2 \, \partial_\beta \ln P_0$ and $\beta^2 \, \partial_\beta P_0$, respectively, while the best fit for $\chi$ as in Eq.~\eqref{eq:chifit} is displayed in yellow.}
    \label{fig:extra-h}
\end{figure}

Since we know from the asymptotic formula for $P_0$ that $\beta^2 \, \partial_\beta \ln P_0$ diverges as $L\to\infty$ let us consider inverse heights. The inverse peak height of $\beta^2 \, \partial_\beta P_0$ tends to $(h_\text{max}^\text{ana})^{-1}\approx0.3009$. If a naive linear fit is applied to $h^{-1}$ as a function of $\ln^{-2} L$ the extrapolation yields $(h_\text{max}^\text{fit})^{-1}\approx {-0.0095(16)}$ for $\beta^2 \, \partial_\beta \ln P_0$ and $(h_\text{max}^\text{fit})^{-1}\approx 0.2161(17)$ for $\beta^2 \, \partial_\beta P_0$. Adding finite-size corrections to the conjectured form of $\chi$ in Eq.~\eqref{eq:chicon} gives~\cite{WJ_05}
\begin{align}\label{eq:chifit}
	 h_{\text{max}}(L) = A_\chi \, L \, \ln^2 L \left( \frac{B_\chi}{\ln L}+\frac{C_\chi}{\ln^2 L}+\frac{D_\chi}{\ln^3 L} \right)
\end{align}
for the maximum of the susceptibility. The peak heights of the three observables and corresponding best fits are shown as function of system size in Fig.~\ref{fig:extra-h}.

\subsection{Peak width~$w$}
\begin{figure}[h]
	\centering
	\epsfig{file=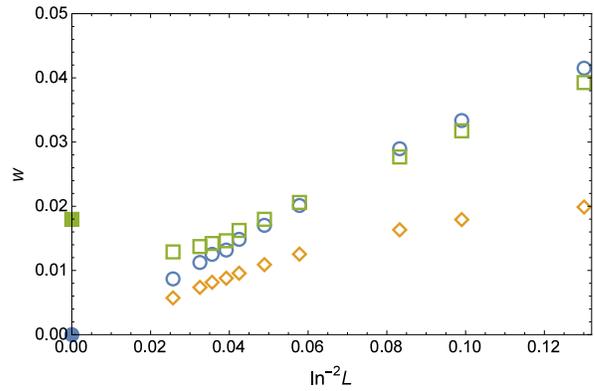,width=0.9\linewidth,clip=}
	\caption{The width, defined as the distance between the peak and the (lower-temperature) position at which the curve reaches $95\%$ of the maximal height, shown for $\beta^2 \, \partial_\beta \ln P_0$ (blue circles), $\beta^2 \, \partial_\beta P_0$ (green squares), and $\chi$ (yellow diamonds) at various system sizes. The asymptotic values, $w^\text{ana}=0$ for $\beta^2 \, \partial_\beta \ln P_0$ and $w^\text{ana}\approx0.0180$ for $\beta^2 \, \partial_\beta P_0$, are indicated at $\ln^{-2} L=0$. Note that in the observed regime all observables decrease monotonically with $L$, yet $\beta^2 \, \partial_\beta P_0$ must increase at some point to reach its asymptotic value.}
    \label{fig:extra-w}
\end{figure}

From the asymptotic expression we know that $w^\text{ana}=0$ for $\beta^2 \, \partial_\beta \ln P_0$ and $w^\text{ana}\approx0.0180$ for $\beta^2 \, \partial_\beta P_0$ in the thermodynamic limit. Our data, together with these analytic values, are shown in Figure~\ref{fig:extra-w}. Since the analytic form of the scaling behaviour for $w$ is lacking no best fit is performed.

\section{Comparison of observables} \label{sec:sec5}

\noindent Using our results we can compare the performance of our new observable for the \textit{F}-model, $\beta^2 \, \partial_\beta \ln P_0$, with that of $\beta^2 \, \partial_\beta P_0$ and that of $\chi$. Asymptotic analytical and numerically extrapolated values for the three characteristics of these observables are collected in Tab.~\ref{tab:table-values} if available.
\begin{table}
	\begin{tabularx}{0.9\linewidth}{*2X *3c} \toprule
		 & & $\beta^2 \, \partial_\beta \ln P_0$ & $\beta^2 \, \partial_\beta P_0$ & $\chi$ \\ \midrule
		\multirow{2}{*}{$\beta_\text{max}$ } & ana & $\ln 2 \approx 0.6931\phantom{(28)}$ & $0.7394\phantom{(17)}$ & $\ln 2$ (conj) \\
		 & fit & $\mspace{51mu} 0.6914(28)$ & $0.6955(17)$ & \ $0.6937(11)$ \ \\ \cmidrule{2-5}
		\multirow{2}{*}{$h_\text{max}^{-1}$ } & ana & $0\mspace{22mu}$ & $0.3009\phantom{(17)}$ & $0$ (conj) \\
		 & fit & $\mspace{37mu}{-0.0095(16)}$ & $0.2161(17)$ & $0$ \\ \cmidrule{2-5}
		\multirow{2}{*}{$w$} & ana & $0\mspace{22mu}$ & $0.0180\phantom{(17)}$ & $0$ (conj) \\
		 & fit & -$\mspace{22mu}$ & -$\mspace{22mu}$ & - \\
		\bottomrule
		\hline
	\end{tabularx}
	\caption{All analytically known and conjectured asymptotic values of our characteristics, together with our numerically extrapolated best values, are shown for our three observables.}
    \label{tab:table-values}
\end{table}

\subsection{Logarithmic derivative of $P_0$}
Our claim is that for an IOPT the logarithmic derivative of the order parameter is a suitable observable for numerical analysis: it must, by construction, tend to a Dirac delta-like distribution at the critical point in the thermodynamic limit. The extrapolated characteristics $\beta^\text{fit}_\text{c}$ and  $h^\text{fit}_\text{max}=-0.0095(16)$ for $\beta^2 \, \partial_\beta \ln P_0$ are in agreement with this claim. Note that a linear fit for the inverse peak height as a function of $\ln^{-2}L$ yields a negative asymptotic result, albeit close to zero, which indicate that there must be other leading finite-size corrections that become important for system sizes outside the reach of the simulations performed here.

\subsection{Ordinary derivative of $P_0$}
It is instructive to compare our new observable with a similar observable that, by construction, should not be suitable for numerical analysis. Interestingly, when the temperature at which $\beta^2 \, \partial_\beta P_0$ peaks is extrapolated in a similar fashion as for the logarithmic derivative the results are comparable. By construction, however, we know that $\beta_\text{max}$ must go to a much higher value in the thermodynamic limit; there must be an inflection point outside the range of simulated system sizes. Similarly, a linear extrapolation for the inverse peak height matches with the data, yet is far from the known asymptotic expression. Concerning the peak width one notes that the observed peaks for $L\geq128$ are less wide than the peak of the asymptotic expression, cf.~the central panel in Fig.~\ref{fig:comparisonplot}; thus $w$ must start to increase at some larger system size, even though it decreases monotonically in the simulated regime.

\subsection{Polarizability}
Finally we turn to $\chi$. Recall that this quantity is not known analytically but there is a conjecture, Eq.~\eqref{eq:chifit}, for its scaling behaviour. The observed $\beta_\text{max}$ for $\chi$ are very close to those of $\beta^2 \, \partial_\beta \ln P_0$, cf.~Fig.~\ref{fig:extra-bcrit}, and the extrapolated value $\beta^\text{fit}_\text{max}=0.6937(11)$ is in agreement with $\beta_c=\ln 2$. Together with the steadily decreasing width for growing system sizes the data suggests that $\chi$ also tends to a Dirac delta-like distribution. Our data fits well with the conjectured form if higher-order finite-size corrections are taken into account, although it must be noted that many alternative forms are also consistent with the data for systems of sizes investigated here.

\section{Conclusion}\label{sec:sec6}

\noindent In this work we looked at infinite-order phase transitions (IOPTs), with the case of the \textit{F}-model as a guiding example. We have suggested a new observable that can be used for finite-size scaling analyses. For any system exhibiting an IOPT with a smooth order parameter this observable is basically the logarithmic derivative of the order parameter, which by construction diverges in the thermodynamic limit. For the \textit{F}-model this is $\beta^2 \, \partial_\beta \ln P_0$, where $P_0$ is the spontaneous staggered polarization. Since the exact asymptotic form of $P_0$ is known in the thermodynamic limit the \textit{F}-model is a good test case to study the performance of our new observable in a finite-size scaling analysis.

For comparison we also have analysed two other observables. The first is $\beta^2 \, \partial_\beta P_0$, which we know to be bounded with peak away from the critical point for all system sizes. Although it must therefore behave quite differently when $L\to\infty$, its observed characteristics turned out to be rather similar to that of $\beta^2 \, \partial_\beta \ln P_0$ at the system sizes investigated. This illustrates that seemingly reasonable yet incorrect conclusions, cf.~the extrapolation to the critical point in Fig.~\ref{fig:extra-bcrit}, may be reached for an IOPT when no analytical expressions are available. The logarithmic corrections and large finite-size corrections for the \textit{F}-model require utmost caution in finite-size analysis; in particular one has to take care to select appropriate observables in order to make hard claims by means of extrapolation to the thermodynamic limit. Given the similarities in FSS of different observables our work thus suggests choosing an observable that is guaranteed to diverge at the critical point. In this way we ensure that the FSS analysis is formally correct, although system sizes large enough to reveal all leading-order corrections will likely be hard to reach.

The final observable that we have investigated is the (spontaneous staggered) susceptibility $\chi=\beta \, [\langle P_0(C)^2 \rangle - \langle P_0(C) \rangle^2]$, which is widely used to analyse phase transitions. The observed characteristics show striking similarities with those of $\beta^2 \, \partial_\beta \ln P_0$ and suggests that $\chi$ also diverges in the thermodynamic limit. The data are compatible with Baxter's conjecture for $\chi$'s scaling behaviour near criticality.

Due to the ice rule the $F$-model is sensitive to the choice of boundary conditions~\cite{KZ_00,*Zin_02}. Certain choices for fixed boundary conditions have already been subjected to some numerical investigations~\cite{AR_05u, TRK_15a,*TRK_15b}. In the near future we intend to analyse the influence of boundary conditions using finite-size scaling. More generally it would be interesting to test our observable for other models with an IOPT such as the \textit{XY}-model.

\section{Acknowledgements}\label{sec:sec7}

\noindent We thank Henk van Beijeren, Henk Blöte, and Henk Stoof for insightful discussions. This work is part of the D-ITP consortium, a program of the Netherlands Organisation for Scientific Research (NWO) that is funded by the Dutch Ministry of Education, Culture and Science (OCW), and is in part funded by the Stichting voor Fundamenteel Onderzoek der Materie (FOM). J.L.~is supported by NWO under the VICI grant 680-47-602.

\bibliography{references}

\end{document}